\documentclass[twocolumn]{IEEEtran}
\usepackage{orcidlink}
\usepackage{url}
\usepackage{indentfirst}
\usepackage{cite}
\usepackage{booktabs}
\usepackage{amsmath,amssymb}
\usepackage{graphicx}
\usepackage{subfigure}
\usepackage{multirow}

\title{Enabling Temporal-Spectral Decoding in Pre-movement Detection}

\author{Hao Jia\orcidlink{0000-0003-1356-7463},
        Cesar F. Caiafa\orcidlink{0000-0001-5437-6095}, 
        Feng Duan~\IEEEmembership{Member,~IEEE}\orcidlink{0000-0002-2179-2460},
        Yu Zhang\orcidlink{0000-0003-4087-6544},
        Zhe Sun\orcidlink{0000-0002-6531-0769}, 
        Jordi Sol{\'e}-Casals\orcidlink{0000-0002-6534-1979}
        \thanks{This work was carried out as part of the doctoral programme in Experimental Science and Technology at the University of Vic - Central University of Catalonia.}
        \thanks{Hao Jia is the Ph.D. student in the Data and Signal Processing Research Group, University of Vic-Central University of Catalonia, Vic, Catalonia.}
        \thanks{Cesar F. Caiafa is working in the Instituto Argentino de Radioastronom\'{i}a, CONICET CCT La Plata/CIC-PBA/UNLP, V. Elisa, Argentina, and Visiting Researcher in the College of Artificial Intelligence, Nankai University, Tianjin, China.}
        \thanks{Feng Duan is working in the College of Artificial Intelligence, Nankai University, Tianjin, China (email: duanf@nankai.edu.cn).}
        \thanks{Yu Zhang is working in the Department of Bioengineering and the Department of Electrical and Computer Engineering, Lehigh University, Bethlehem, PA 18015, USA.}
        \thanks{Zhe Sun is working in the Computational Engineering Applications Unit, Head Office for Information Systems and Cybersecurity, RIKEN, Saitama, Japan (email: zhe.sun.vk@riken.jp).}
        \thanks{Jordi Sol{\'e}-Casals is working in the Data and Signal Processing Research Group, University of Vic - Central University of Catalonia, Vic, Catalonia, and Visiting Researcher in the Department of Psychiatry, University of Cambridge, United Kingdom and in the College of Artificial Intelligence, Nankai University, Tianjin, China (email: jordi.sole@uvic.cat).}}

\begin{document}
\maketitle
\begin{abstract}
Non-invasive brain-computer interfaces help the subjects to control external devices by brain intentions. The multi-class classification of upper limb movements can provide external devices with more control commands. 
The onsets of the upper limb movements are located by the external limb trajectory to eliminate the delay and bias among the trials. However, the trajectories are not recorded due to the limitation of experiments. The delay cannot be avoided in the analysis of signals.
The delay negatively influences the classification performance, which limits the further application of upper limb movements in the brain-computer interface.

This work focuses on multi-channel brain signals analysis in the temporal-frequency approach. It proposes the two-stage-training temporal-spectral neural network (TTSNet) to decode patterns from brain signals. The TTSNet first divides the signals into various filter banks. In each filter bank, task-related component analysis is used to reduce the dimension and reject the noise of the brain. A convolutional neural network (CNN) is then used to optimize the temporal characteristic of signals and extract class-related features. Finally, these class-related features from all filter banks are fused by concatenation and classified by the fully connected layer of the CNN.


The proposed method is evaluated in two public datasets. The results show that TTSNet has an improved accuracy of 0.7456$\pm$0.1205 compared to the EEGNet of 0.6506$\pm$0.1275 ($p<0.05$) and FBTRCA of 0.6787$\pm$0.1260 ($p<0.1$) in the movement detection task, which classifies the movement state and the resting state. The proposed method is expected to help detect limb movements and assist in the rehabilitation of stroke patients.
\end{abstract}


\section{Introduction}
\label{sec:intro}

Non-invasive brain-computer interface bridges the gap between the human brain and external devices like computers and robots \cite{schalk_practical_2010,hramov_physical_2021,saha_progress_2021,kawala-sterniuk_summary_2021}. It converts brain activities into control commands by analyzing the electroencephalogram (EEG) information. The EEG signals are acquired from the brain scalp with acquisition devices in a non-invasive approach. 

Two widely learned brain activities can provide stable control commands, motor imagery and steady-state visual evoked potential \cite{zhang_motor_2020,zhang_temporally_2019,mattioli_1d_2021,jiao2018novel,nakanishi2018enhancing,liu_improving_2021}. Motor imagery is usually related to the left/right limb movement. When the limb on one side moves, the power in the motor cortex changes on the contra side \cite{zhang_temporally_2019}. The steady-state visual evoked potential is evoked by the external visual stimuli when the subjects focus their eyes on the stimuli. The EEG signals from the visual cortex have the same frequency as the external visual stimuli \cite{nakanishi2018enhancing}. Instead of the well-studied left/right limb movements or the visual-evoked potential, this work focuses on the actual movements of the one-side upper limb. 

Movement-related cortical potential (MRCP) is a brain activity related to the upper limb movement \cite{olsen_paired_2018,ofner_upper_2017,ofner_attempted_2019,farina_introduction_2013}. When the subject's limb moves, the EEG signals acquired from the motor cortex have an amplitude increase in the low-frequency domain \cite{ofner_upper_2017,ofner_attempted_2019}. Because the EEG signals are influenced by noises during acquisition, the signals are averaged across the trials of the repeated motions and thus reduce the influence of noises. This averaged signal is the grand average MRCP \cite{borras_influence_2022}. When the limb executes different motions like hand close and elbow flexion, the grand average MRCPs are different before and after the movement onset. The differences of these grand average MRCPs are used to discriminate the upper limb movements of subjects \cite{ofner_upper_2017}.

The previous studies on MRCP signals have two tasks, movement detection and movement classification \cite{kaeseler_feature_2022}. The movement detection is to distinguish movement from the resting state, which is a binary classification between the movement state and the resting state. The movement classification is the binary or multi-class classification of movement states. In the movement states, when the limb moves, the grand average MRCP fluctuates and has an increase followed by a decrease around the movement onset. The grand average MRCP in the resting state is relatively steady compared to the movement state \cite{ofner_upper_2017}. The movement detection, classified between the fluctuating and the steady signals, has a higher classification accuracy compared to the movement classification. Movement detection is a subset of movement classification, but movement classification is still under development.

The method of subject-dependent and section-wise spectral filtering solves the movement detection problem by considering the MRCP signals for two different temporal sections \cite{jeong_decoding_2020}. The two sections are a two-second time window before the movement onset and a one-second time window after the movement onset. The signals in the two sections are averaged. The mean amplitudes in both sections are used as the features and are fed into the regularized linear discriminant analysis classifier. This method uses the changes in amplitude of MRCP signals before and after the movement onset. However, the averaging of the signals ignores how the amplitude changes and leads to information loss in signal processing.

In our previous work, we proposed the standard task-related component analysis method (STRCA) to solve the movement detection problem \cite{duan_decoding_2021}. This method uses task-related component analysis as the spatial filter to remove the noises and task-unrelated components from EEG signals. After applying spatial filtering to the unlabelled signals and the grand average MRCPs, the canonical correlation coefficients are used to measure the similarity between the unlabelled signals and the multiple grand average MRCPs. The linear discriminant analysis classifies the extracted coefficients and predicts the label of signals. To use the information about how the amplitude changes in MRCP signals, STRCA compares the unlabelled signal and the grand average MRCPs and uses the similarity as the features.

Frequency domain analysis is a universal approach to analyzing time series. To use the information in different frequencies, we proposed the filter bank task-related component analysis method (FBTRCA) \cite{jia_improve_2022}. The EEG signals are divided into various filter banks. These filter banks have the same low cut-off frequency at 0.5 $Hz$, and the high cut-off frequencies are sorted in the arithmetic sequence from 1 $Hz$ to 10 $Hz$. The STRCA is used to extract features from each filter bank, and all these features are concatenated. The feature selection method, based on mutual information, minimum redundancy maximum relevance, is used to select the best features from the set. The essential features are finally classified by the support vector machine classifier.

In STRCA, the spatial filter is obtained by concatenating the task-related component analysis spatial filters of two classes. When the number of classes increases, the number of vectors along the axis of concatenation is greater than the number of EEG channels. The concatenation limits the usage of STRCA and FBTRCA in multi-class tasks. To be able to use STRCA and FBTRCA in the multi-class classification, we optimized the structure of the spatial filter and removed the common component from the grand average MRCP. The FBTRCA can be used in both movement detection, and movement classification \cite{jia_toward_2022}.

The signal analysis on the time domain helps to better explain the EEG signals. For example, in motor imagery, the utilization of sliding time windows improves the classification performance between the left and right limbs. The deep learning method EEGNet is an efficient tool in dealing with EEG signals \cite{lawhern_eegnet_2018}. EEGNet can use the temporal information in EEG signals because of its shift-invariant in the convolutional layers. 

Our previous proposed method, FBTRCA, uses the information in various filter banks. The temporal information is measured simply by the correlation of unlabelled EEG signals and the grand average MRCPs. The correlation highly relies on the grand average MRCPs. However, the noise in the grand average MRCP is simply rejected by averaging the signals belonging to the same class. It means that the grand average MRCPs used in FBTRCA may be influenced by unknown noises. Besides, when the movement onset is unknown, the latency lag may exist between the MRCP signals of two trials \cite{sburlea_continuous_2015}. In this case, shift invariance is necessary for the signal processing of the MRCP signal. The FBTRCA cannot use the temporal information in EEG signals efficiently. 

In this work, we aim to further improve the performance of FBTRCA when the movement onset cannot be located. When this happens, the movement onset is unknown, and the latency among the movement onset of trials has a negative influence on the classification performance. This work proposes two-stage-training temporal-spectral neural network (TTSNet), which considers the shift-invariant of convolutional layers and incorporates temporal information into the FBTRCA method. TTSNet has four steps, (1) dividing EEG signals into filter banks, (2) optimizing EEG signals with the spatial filter, (3) extracting temporal features with EEGNet, and (4) concatenating features and classifying the features with a fully connected layer.

The structure of this work is organized as follows. Section \ref{sec:method} introduces the used EEG datasets and how the datasets are preprocessed. Besides, the structure of the proposed method is given. Section \ref{sec:result} evaluates the classification performance in the EEG dataset. The proposed method is compared against the other state-of-the-art methods. Section \ref{sec:discuss} explains how the proposed method is designed and how the proposed method uses information from the grand average MRCP. Finally, Section \ref{sec:conclu} concludes this paper.

\section{Method}
\label{sec:method}

\subsection{Dataset Description}
\label{ssec:dataset}
In this work, two public datasets about upper limb movement, Dataset I and Dataset II, are used to evaluate the classification performance. Dataset I consists of EEG data from 15 healthy subjects \cite{ofner_upper_2017}. Dataset II consists of EEG data from 10 subjects with cervical spinal cord injuries \cite{ofner_attempted_2019}. The preprocessing procedure of the raw EEG signals follows the same operation in our previous works, which is given in the following. 

Each of the subjects was asked to sit in front of a computer, and the arm was supported by a table or exoskeleton to avoid muscle fatigue. The signal acquisition paradigm is trial-based. A trial lasts 5 seconds. At 0 $s$, the trial starts with a beep, and a cross is displayed on the computer screen. Two seconds later, a cue was presented on the computer screen. The subjects were expected to execute the required movement related to the cue.

The movements executed in dataset I include \textit{elbow flexion}, \textit{elbow extension}, \textit{supination}, \textit{pronation}, \textit{hand close}, \textit{hand open}, and \textit{resting}. In dataset II, the executed movements include \textit{supination}, \textit{pronation}, \textit{hand open}, \textit{palmar grasp}, and \textit{lateral grasp}. The numbers of trials per class are 60 and 72 in dataset I and dataset II, respectively.

In dataset I, the movement trajectories of the hand were recorded by the exoskeleton simultaneously during the acquisition of EEG signals. The movement onset when the limb actually moves can be located by the recorded movement trajectory. However, in dataset II the movement trajectories were not recorded, so the movement onset for each trial is unknown. The latency lag between the movement onset and the cue cannot be eliminated in dataset II. The proposed TTSNet aims to reduce the influence of the latency lag on classification performance. To evaluate the performance of TTSNet, the experiments of dataset I are divided into two cases, dataset I(a) and dataset I(b). In dataset I(a), the movement trajectories of dataset I are used to locate the movement onset. The EEG signals from two seconds before the movement onset to one second after the movement onset are used in the evaluation and classification task. In dataset I(b), it is assumed that the movement trajectories were not recorded, so the contaminated trials cannot be rejected, and the movement onset cannot be located. The EEG signals from one second before the cue to two seconds after the cue are used in the classification task. In dataset II, the range of EEG signals for classification is a two-second time window from the cue.

Because dataset I has the simultaneously acquired hand trajectories, but dataset II does not, the movement onset can be located with the trajectories in dataset I. In the localization of the movement onset in dataset I, the 1-order difference of the trajectory is first smoothed by the 1-order \textit{Savitzky-Golay} finite impulse response smoothing filter with time window length 31. The filtered 1-order difference is then normalized by the maximum absolute value. For trials belonging to the \textit{resting} state, a fake movement onset is set to 2.5 $s$ after the trial starts. Trials in the \textit{resting} state are rejected if the variances of normalized trajectory are less than 0.02. In \textit{elbow flexion} and \textit{elbow extension}, the amplitude of the hand trajectory is higher than the other four motions because the limb moves. The movement onset is set to the location where the normalized trajectory equals the threshold of 0.05. Trials are rejected manually when the movement onset is highly influenced by noise contamination. In the other four motions, the function $f(x)=a*exp(-(\frac{x-b}{c})^2)+d$ is used to fit the smoothed and normalized trajectory by tuning the parameters $a,b,c,d$ \cite{duan_decoding_2021,jia_improve_2022}. Trials are rejected if the parameters of the tuned function fulfill $a<0.05$, $c>100$, and $d>10$. The movement onset is set to the time point whose amplitude equals 0.1.

The EEG signals for classification were acquired from the motor cortex of the brain, namely $FC_z$, $C_3$, $C_z$, $C_4$, $CP_z$, $F_3$, $F_z$, $F_4$, $P_3$, $P_z$ and $P_4$. The raw EEG signals are firstly downsampled to 256 $Hz$. The z-normalization is then used to normalize the EEG signals. Because the MRCP signals are located at the low-frequency bands of EEG signals, the normalized EEG signals are bandpassed to 0.5$\sim$10 $Hz$.

\subsection{Two-stage-training Temporal-Spectral Network}
\label{ssec:ttsnet}
\begin{figure*}[htbp]
    \centering
    \includegraphics[width=\textwidth]{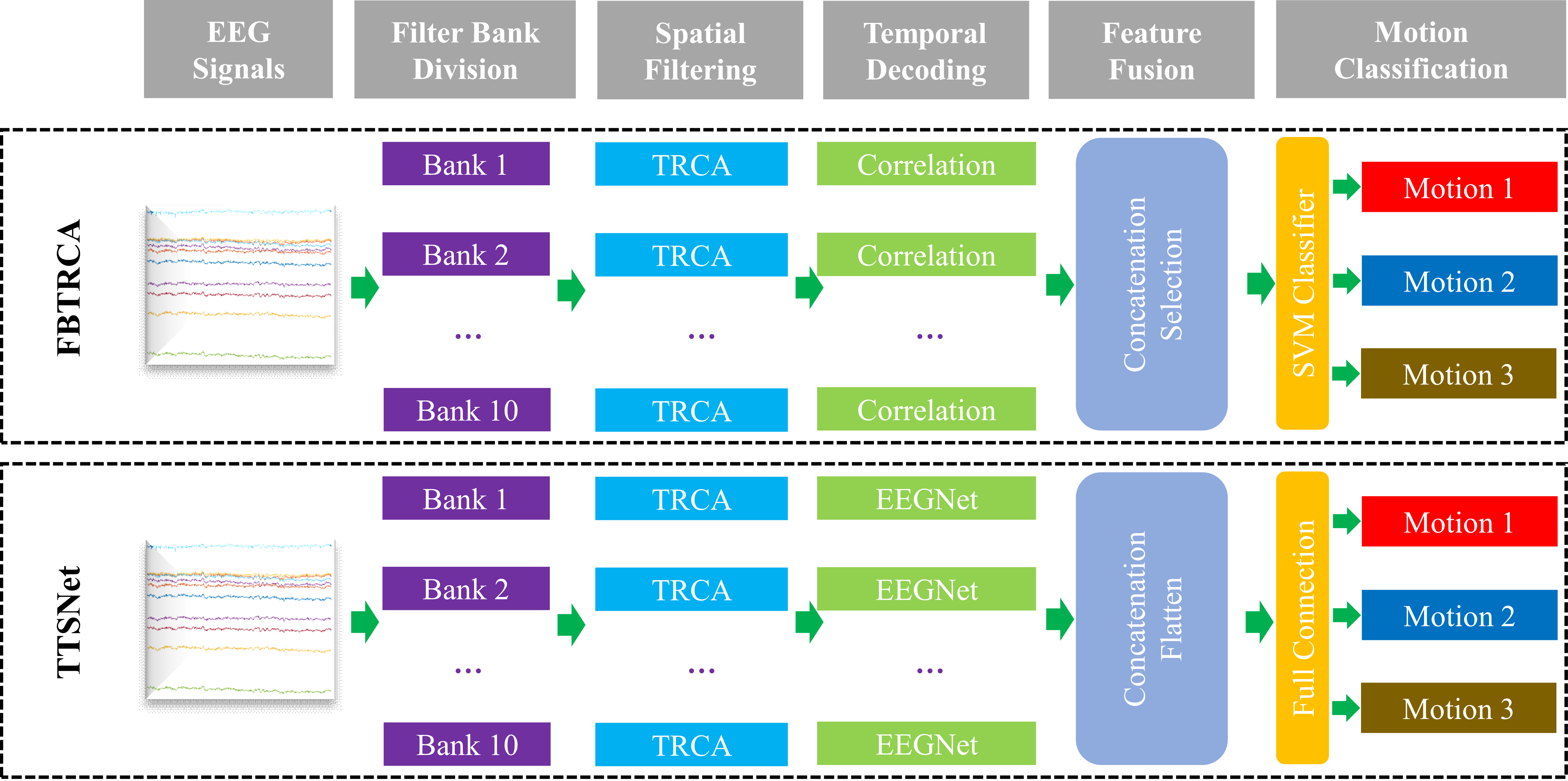}
    \caption{The overall framework of the multi-class FBTRCA method and the TTSNet method. In both methods, EEG signals are divided into filter banks and then optimized with the spatial filter TRCA. During decoding the temporal information from signals, FBTRCA uses the correlation between these signals and the grand average MRCPs as the features. TTSNet uses the EEGNet to capture the temporal information and use the shift-invariant of the convolution layers. After decoding the temporal information, output features from all the filter banks are concatenated. In FBTRCA, these features are selected and optimized by the minimum redundancy maximum relevance method and classified by the support vector machine (SVM) method. In TTSNet, these features are flattened and then classified by the fully connected layer.}
    \label{fig:2.1}
\end{figure*}

The two-stage-training temporal-spectral network (TTSNet) is further developed based on the FBTRCA method and has four steps: (1) filter bank division, (2) spatial filtering, (3) temporal decoding, and (4) feature fusion and classification. The overall framework of both the FBTRCA and TTSNet is given in Fig. \ref{fig:2.1}, which shows the relationship and the differences between the two methods.

\subsubsection{Filter Bank Division}
\label{sssec:bankdivide}
The MRCP signals are located at the low-frequency bands of EEG signals. The approximate range of the low-frequency bands is 0.5$\sim$10 $Hz$. To use the information in the frequency domain of MRCP signals, we proposed a filter bank division method for the low-frequency bands. The low cut-offs of these bands are fixed at 0.5 $Hz$. The
high cut-offs of these bands are sorted as an arithmetic sequence from 1 $Hz$ to 10 $Hz$ with step 1 $Hz$. Therefore, the range of the low-frequency bands, 0.5$\sim$10 $Hz$, is divided into $F$ filter banks, where $F=10$.

\subsubsection{Spatial Filtering}
\label{sssec:spatial}

After the filter bank division, the MRCP signals are divided into various filter banks. In each filter bank, the multi-channel signals contain task-unrelated components because the filter bank division cannot remove the noises from the original signals. Here, we use the spatial filter to reject the noises and remove task-unrelated components from the original signals in each filter bank. A spatial filter is a matrix $W$ with size $C\times P$, where $C$ is the number of channels and $P$ is an integer smaller than $C$. Spatial filtering is the operation that multiplies the spatial filter and the raw EEG signals. By the matrix multiplication between the original signals $X\in \mathbb{R}^{C\times T}$ and the $W$, the spatial-filtered signals $X^TW\in \mathbb{R}^{T\times P}$ has a decreased dimension, and the noises are removed. Here, task-related component analysis (TRCA) is used as the spatial filter. TRCA aims to find a $W$ that maximizes the inter-trial covariances within a class. The training set belonging to the class $k$ is $\mathcal{X}^k=\{X_1, X_2, ..., X_N\}$, where $N$ is the number of trials of class $k$ and $X_N\in\mathbb{R}^{C\times T}$. The inter-trial covariance of class $k$ is computed with the equation:
\begin{equation}
    S^k=\sum_{i,j=1,i<j}^{N}X_i^TX_j+X_j^TX_i.
\end{equation}

To normalize the inter-trial covariance, the self-covariance is introduced:
\begin{equation}
    Q^k=\sum_{i=1}^{N}X_i^TX_i.
\end{equation}

The spatial filter TRCA is obtained by solving the Eigen equation:
\begin{equation}
    \max_{\omega}{J} = \frac{\omega^TS\omega}{\omega^TQ\omega}
\end{equation}
where $\omega\in\mathcal{R}^{C\times 1}$ is the eigenvector. $S$ and $Q$ are the matrices that summarized the inter-covariances and self-covariances of $K$ classes:
\begin{equation}
    S = \sum_{k=1}^{K} S^k, Q = \sum_{k=1}^{K} Q^k.
\end{equation}

The eigenvectors $\omega$ of maximum eigenvalues are concatenated into the spatial filter $W\in\mathbb{R}^{C\times P}$, where $P$ is the number of selected eigenvectors.

\subsubsection{Temporal Decoding}
\label{sssec:temporal}

Temporal decoding plays the role of feature extraction in both the FBTRCA and the TTSNet. In FBTRCA, the correlated coefficient is used as the feature that measures the differences between the unlabelled EEG signals and the grand average MRCPs. However, the correlation coefficient can only capture stationary temporal characteristics and cannot deal with the shifted temporal characteristic. Therefore, we propose the TTSNet by improving the temporal decoding ability of FBTRCA with the convolutional neural network.

\begin{figure}[htbp]
    \centering
    \includegraphics[width=0.475\textwidth]{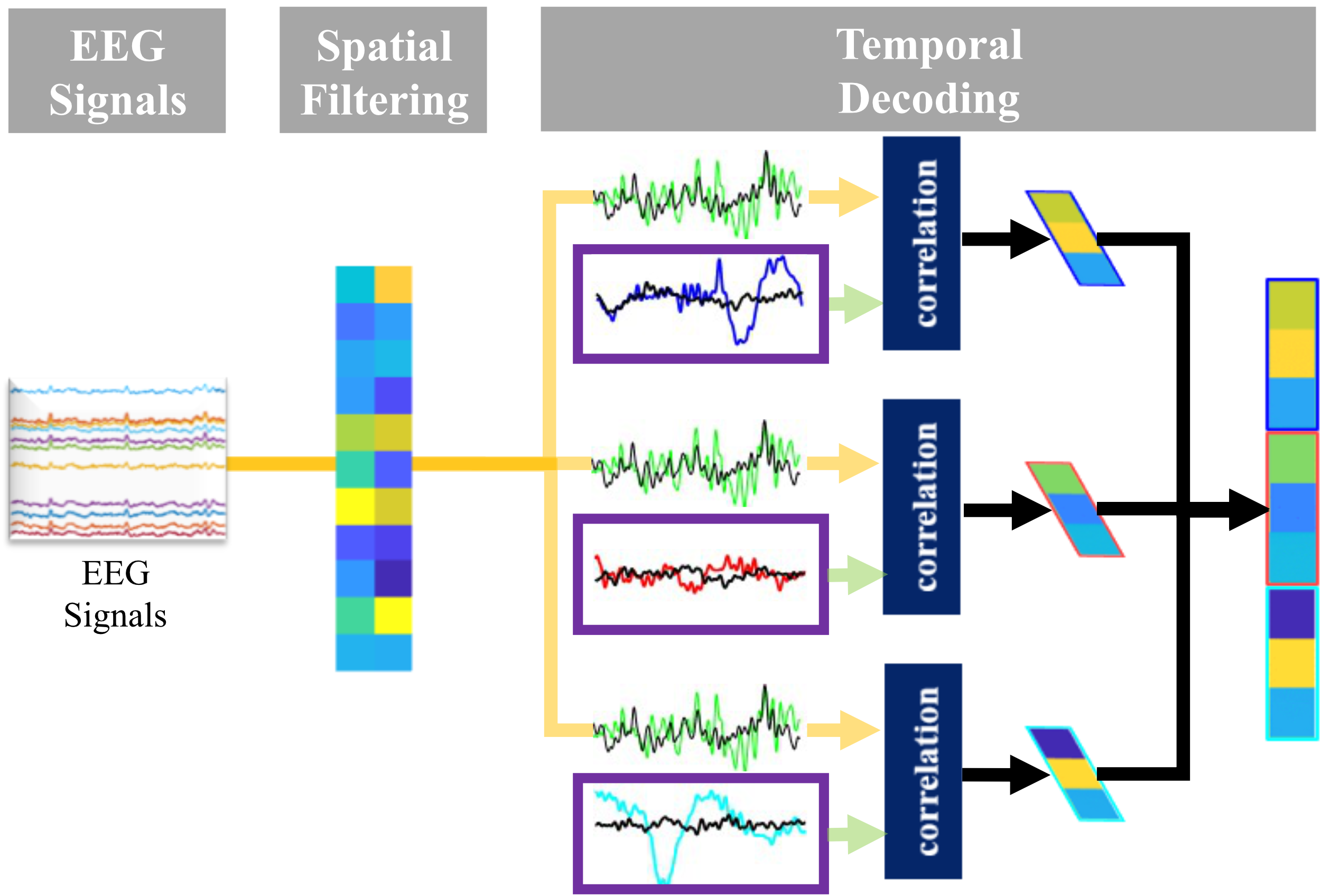}
    \caption{The structure of temporal decoding part from one filter bank in the FBTRCA method. The EEG signals are first filtered with the spatial filter TRCA to remove the unrelated noises and reduce the dimensions of signals. To use the information about the amplitude changes of filtered signals, the correlation between the filtered signals and the filtered grand average MRCPs (the purple box) of these classes are computed as the features. These correlation features of all the classes are concatenated as the output features of this filter bank.}
    \label{fig:2.2}
\end{figure}

\paragraph{Correlation Coefficient}
\label{ssssec:corrcoef}
The correlation coefficient measures the similarity of two matrices. In FBTRCA, the two input matrices are the spatial-filtered unlabelled signals and the spatial-filtered grand average MRCPs. In Fig. \ref{fig:2.2}, the structure of computing correlation coefficients in FBTRCA is given.

The grand average MRCP of class $k$ is obtained from the training set $\mathcal{X}^k$ by taking the average of all trials:
\begin{equation}
    \hat{X^k}=\frac{1}{N}\sum_{i=1}^{N}X_i
\end{equation}

The common component of the grand average MRCPs of $K$ classes is firstly removed from both the grand average MRCPs $\hat{X^k}$ and the unlabelled signals $X\in\mathbb{R}^{C\times T}$:
\begin{equation}
    \hat{X^k_\&} = \hat{X^k} - \frac{1}{K}\sum_{k=1}^{K}\hat{X^k},
    X_\& = X - \frac{1}{K}\sum_{k=1}^{K}\hat{X^k}
\end{equation}

The correlation coefficient is a normalized point-wise product of two input matrices. Given two input matrices $X, Y\in\mathbb{R}^{I\times J}$ that fulfil $mean(X)=0$ and $mean(Y)=0$, the correlation coefficient is computed using:
\begin{equation}
    r = corr(X, Y)=\frac{X*Y}{\sqrt{(X*X)\times(Y*Y)}}
    \label{eqn:corr}
\end{equation}
where $*$ denotes the summed-up point-wise products of two input matrices. The symbol $\times$ multiplies two constants $X*X$ and $Y*Y$.

Three kinds of correlation coefficients are computed by taking $X_*=X_\&^TW$ and $X_k=\hat{X^k_\&}^TW$ as the inputs.

\noindent(1) Correlation Coefficient
\begin{equation}
    \rho_{1,k}=corr(X_*,X_k);
\end{equation}

\noindent(2) Canonical Correlation Coefficient
\begin{gather}
    [A, B] = cca(X_*, X_k)\\
    \rho_{2,k}=corr(X_*B,X_kB);
\end{gather}

\noindent(3) Normalized Canonical Correlation Coefficient
\begin{gather}
    [A, B] = cca(X_*-X_k, X_{-k}-X_k)\\
    \rho_{3,k}=corr((X_*-X_k)A, (X_{-k}-X_k)A);
\end{gather}
where $X_{-k}$ is the mean of spatial-filtered grand average MRCPs of all classes except for class $K$:
\begin{equation}
    X_{-k} =\frac{1}{K-1}\sum_{kk=1,kk\neq k}^{K}X_{kk}.
\end{equation}

Therefore, there are $3 \times K \times F$ features in FBTRCA, where $F$ is the number of filter banks.

\paragraph{Convolution Neural Network}

The correlation coefficients in FBTRCA measure the similarity between unlabelled trials and the grand average MRCPs and thus decode the temporal characteristics in the signals. However, the temporal decoding of FBTRCA needs to be further improved for two reasons:

\noindent (1) The grand average MRCP is averaged across trials belonging to the same class. However, averaging cannot ensure the center of trials in the training and testing set is learned;

\noindent (2) When the movement onset cannot be located, or the located onset is biased, the MRCP signals are shifted from the real onset. Correlation cannot deal with the shifted onset.

The correlation coefficient in Equation \ref{eqn:corr} has two inputs. One is the input MRCP signals. The other one is the grand average MRCP, which can be regarded as the pre-trained weights from the training set by averaging EEG signals across trials. The grand average MRCP, \textit{i.e.}, the weight applied to the input signals, can be further improved because the weight is obtained very simply by averaging across trials.

The convolution neural network (CNN) has two advantages in replacing the role of correlation in temporal decoding: (1) trainable weights and (2) shift-invariant property. Here, we adopt the network architecture of EEGNet in temporal decoding because of the good performance of EEGNet in classification. The network architecture is given in Table \ref{tab:2}.

\begin{table}[htbp]
    \centering
    \caption{The model structure of temporal decoding in TTSNet}
    \begin{tabular}{c|c|c}
        \toprule
        Layer & Output Size & Parameter\\
        \midrule
        Input Layer & [$B$, 1, $C$, $T$] &\\
        \midrule
        ZeroPad2d & [$B$, 1, $C$, $T$+63] & (31, 32, 0, 0)\\
        Conv2d & [$B$, 8, $C$, $T$] & (1, 64)\\
        BatchNorm2d & [$B$, 8, $C$, $T$] &\\
        Conv2d & [$B$, 16, 1, $T$] & ($C$, 1), $grouped$\\
        BatchNorm2d & [$B$, 16, 1, $T$] &\\
        ELU & [$B$, 16, 1, $T$] &\\
        AvgPool2d & [$B$, 16, 1, $T$//4] & (1, 4)\\
        Dropout & [$B$, 16, 1, $T$//4] & 0.25\\
        \midrule
        ZeroPad2d & [$B$, 16, 1, $T$//4+15] & (7, 8, 0, 0)\\
        Conv2d & [$B$, 16, 1, $T$//4] & (1, 15), $grouped$\\
        Conv2d & [$B$, 16, 1, $T$//4] & (1, 1)\\
        BatchNorm2d & [$B$, 16, 1, $T$//4] &\\
        ELU & [$B$, 16, 1, $T$//4] &\\
        AvgPool2d & [$B$, 16, 1, $T$//32] & (1, 8)\\
        Dropout & [$B$, 16, 1, $T$//32] & 0.25\\
        \midrule
        Flatten & [$B$, 16*$T$//32] &\\
        Linear & [$B$, $K$] & $bias=False$\\
        \bottomrule
    \end{tabular}
    \label{tab:1}
\end{table}

\subsubsection{Classification}
\label{sssec:filterbank}

In FBTRCA, the features of all filter banks are sorted and selected by the minimum redundancy maximum relevance method and then classified by the support vector machine classifier. The proposed TTSNet uses a fully connected layer to optimize these features and classify the features. The architecture of the fully connected layer is given in Table \ref{tab:2}.

\begin{table}[htbp]
    \centering
    \caption{The model structure of classification in TTSNet}
    \begin{tabular}{c|c|c}
        \toprule
        Layer & Output Size & Parameter\\
        \midrule
        Input Layer & [$B$, $K$, $F$] &\\
        \midrule
        Flatten & [$B$, $K*F$] & \\
        Linear & [$B$, $K*F*2$] & $bias=False$\\
        Relu & [$B$, $K*F*2$] &\\
        Linear & [$B$, $K*F//2$] & $bias=False$\\
        Relu & [$B$, $K*F//2$] &\\
        Linear & [$B$, $K$] & $bias=False$\\
        \bottomrule
    \end{tabular}
    \label{tab:2}
\end{table}

\subsubsection{Two-stage Training}
\label{sssec:tsteptrain}
The TTSNet has two modules that consist of neural networks, the CNN for temporal decoding and the fully connected layer for classification. During training the TTSNet, the two modules are trained separately, \textit{i.e.}, in a two-stage approach. In the first stage, a CNN will be trained for each of the filter banks, and thus the number of trained CNNs is $F$. The output of $F$ CNNs is concatenated and flattened. In the second stage, the fully connected layer is trained with the flattened features from CNNs.
As given in Table \ref{tab:1}, the output layer is a linear layer with output size $K$, where $K$ is the number of classes. Therefore, the network can be trained with the losses between the train label and the outputs.
The train label is used twice in the two-stage training process, the temporal decoding in Table \ref{tab:1} and the classification in Table \ref{tab:2}. In the first stage, the weights in the fully connected layer are fixed, and the CNNs are trained with train labels. In the second stage, the weights in the CNNs are fixed, and the fully connected layer is trained with train labels.

\subsection{Comparison Methods}
\label{ssec:sota}

In our previous work \cite{jia_toward_2022}, the FBTRCA method was compared against both machine learning methods and deep learning methods, such as multi-class common spatial pattern \cite{grosse-wentrup_multiclass_2008}, minimum distance to mean \cite{barachant_multiclass_2012}, WaveNet \cite{thuwajit_eegwavenet_2022}, and Deep CNN \cite{schirrmeister_deep_2017}. In this work, we don't further compare these methods and use the FBTRCA as the benchmark. The methods for comparison in this work have
two state-of-the-art methods (FBTRCA, EEGNet) and two baseline methods.

\paragraph{Baseline I: TEGNet} Compared to the EEGNet, The TTSNet fuses the temporal features from all filter banks. To bridge the gap between the TTSNet and the EEGNet in the comparison, we additionally introduce the Tasked-related EEGNet (TEGNet). The TEGNet has the spatial filtering and temporal decoding part of TTSNet and only processes signals from one filter bank.

\paragraph{Baseline II, OTSNet} The TTSNet is trained by two steps. To show the necessity of the two-stage training, the performance of TTSNet is compared to the one-step-training temporal-spectral network (OTSNet), which is training in one step without training the temporal decoding CNN module additionally.

\section{Result Analysis}
\label{sec:result}

\subsection{Parameter Setting}
\label{ssec:parameter}
The classification tasks in this experiment include (1) the binary classification between two motions, \textit{e.g.}, \textit{elbow flexion} and \textit{elbow extension}, and (2) the multi-class classification among all the motions in each dataset. In both classification tasks, the two datasets are split into the training set and the testing set by ten-fold cross-validation. The mean and standard deviation of the accuracy of ten folds are used to evaluate the performance of the mentioned classification methods.

The training process of EEGNet, TEGNet, OTSNet, and TTSNet all has the parameters: learning rate (0.001), batch size (50), optimizer (Adam), and loss function (cross-entropy). In TTSNet, the network is trained in two steps. The CNN module is firstly trained for 200 epochs (in the binary classification) and 50 epochs (in the multi-class classification). The classification accuracy has converged within the number of training epochs. The model weight of the CNN module is then not updated in the following training process of the fully connected layer. In the training process of the second module, the Adam optimizer is additionally equipped with the weight decay (L2 penalty) of 0.1 to ensure the early stop of the training process.

In the FBTRCA, TEGNet, OTSNet, and TTSNet, the spatial filter is necessary during the computation. The number of selected maximum eigenvalues $P$ is a hyper-parameter. According to our previous work on FBTRCA, $P$ is set to 3 when the movement onset is located and aligned for all trials; $P$ is set to 6 when the movement onset cannot be located.

\subsection{Result}
\label{ssec:result}

Five methods are compared in this section, including the proposed TTSNet method and the baseline methods FBTRCA, EEGNet, TEGNet, and OTSNet. EEG signals from Dataset I and Dataset II are used in three cases: (1) Dataset I(a)-Dataset I with the movement onset located, (2) Dataset I(b)-Dataset I without the located movement onset and (3) Dataset II without the located movement onset. 

\begin{figure*}[htbp]
    \centering
    \includegraphics[width=\textwidth]{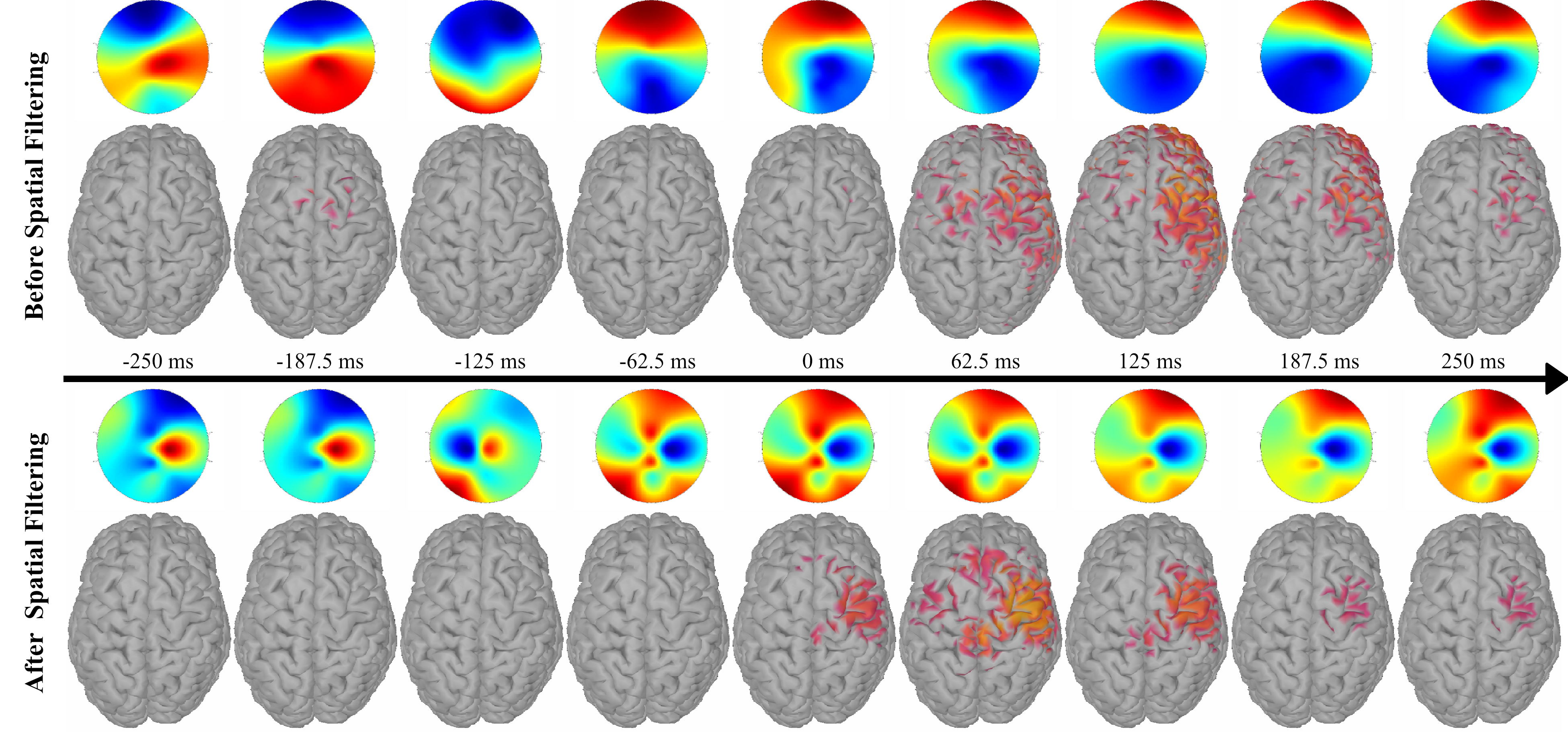}
    \caption{Comparison of the grand average MRCP signals (0.5$\sim$10 $Hz$) before and after the spatial filtering. This figure visualizes the grand average MRCP signals of elbow flexion of subject 1 from dataset I in both the channel space and the source space. The 0 ms in the figure denotes the movement onset. When visualizing the signals after spatial filtering, the problem is that the number of channels of filtered signals is decreased. To solve this problem, the filtered signals $X^TW$ are multiplied by the inverse spatial filter $W^T$, and we get $X^TWW^T$. $X^TWW^T$ reserves the filtered signals to the original channels that are the same as the channels before spatial filtering.}
    \label{fig:3.0}
\end{figure*}

Before the statistic analysis in the three cases, we first compare the EEG signals before and after spatial filtering (with task-related component analysis) in the channel space and the source space, which is given in Fig \ref{fig:3.0}. The figures for the channel space are plotted with EEGLab \cite{delorme_eeglab_2004}. The figures for the source space are plotted with Brainstorm \cite{tadel_brainstorm_2011}. During plotting the source space figures, the brain anatomy used is BCI-DNI BrainSuite \cite{joshi_hybrid_2022}. The surface is segmented by the boundary element method with Brainstorm \cite{frijns_improving_2000}. The source spaces figures are finally given by the standardized low-resolution brain electromagnetic tomography method \cite{pascual_marqui_standardized_2002}.

\subsubsection{Overall Comparison}
\label{ssec:result:all}

\begin{figure*}[!htbp]
    \centering
    \subfigure[Dataset I(a)]{
        \includegraphics[width=0.315\textwidth]{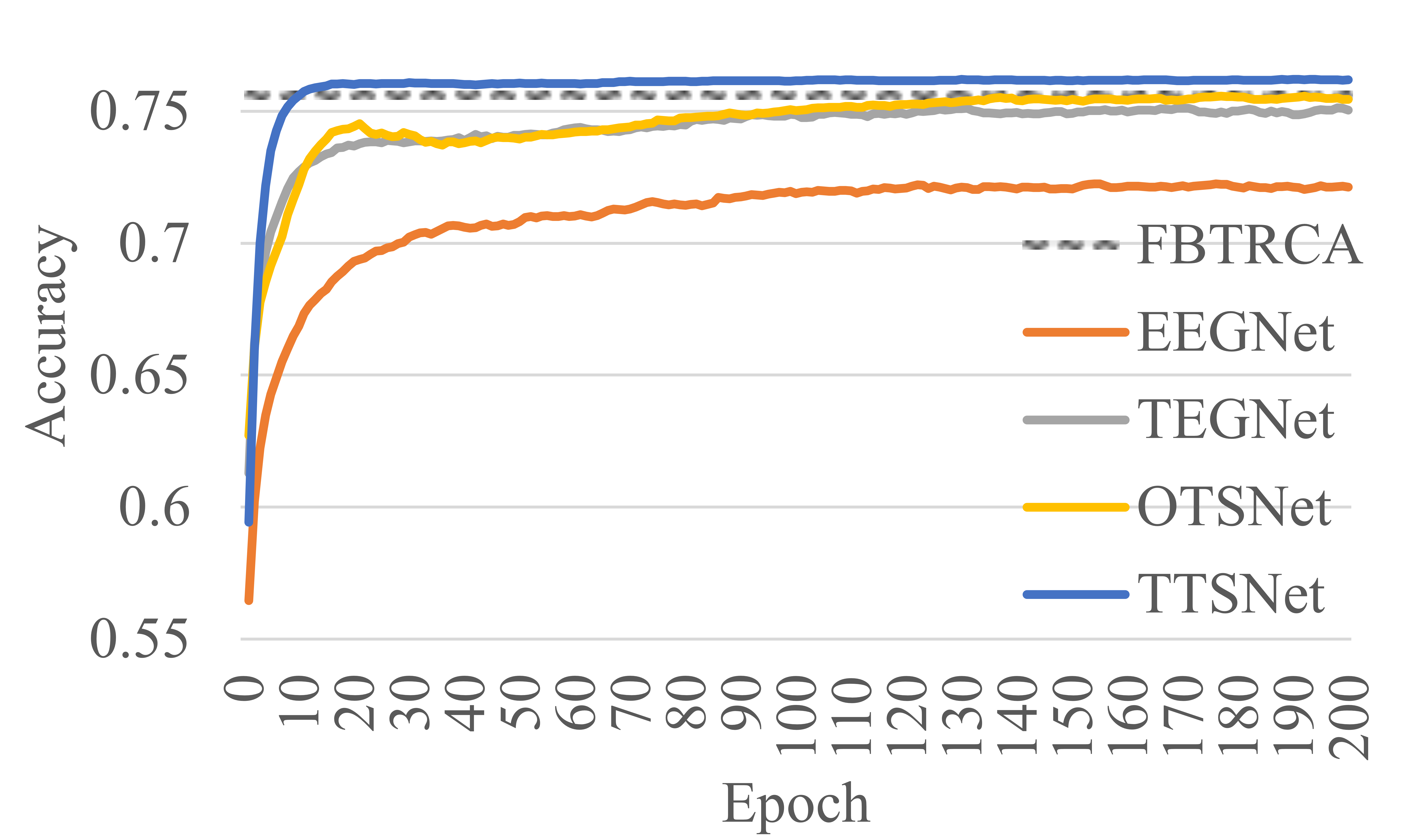}
    }
    \subfigure[Dataset I(b)]{
        \includegraphics[width=0.315\textwidth]{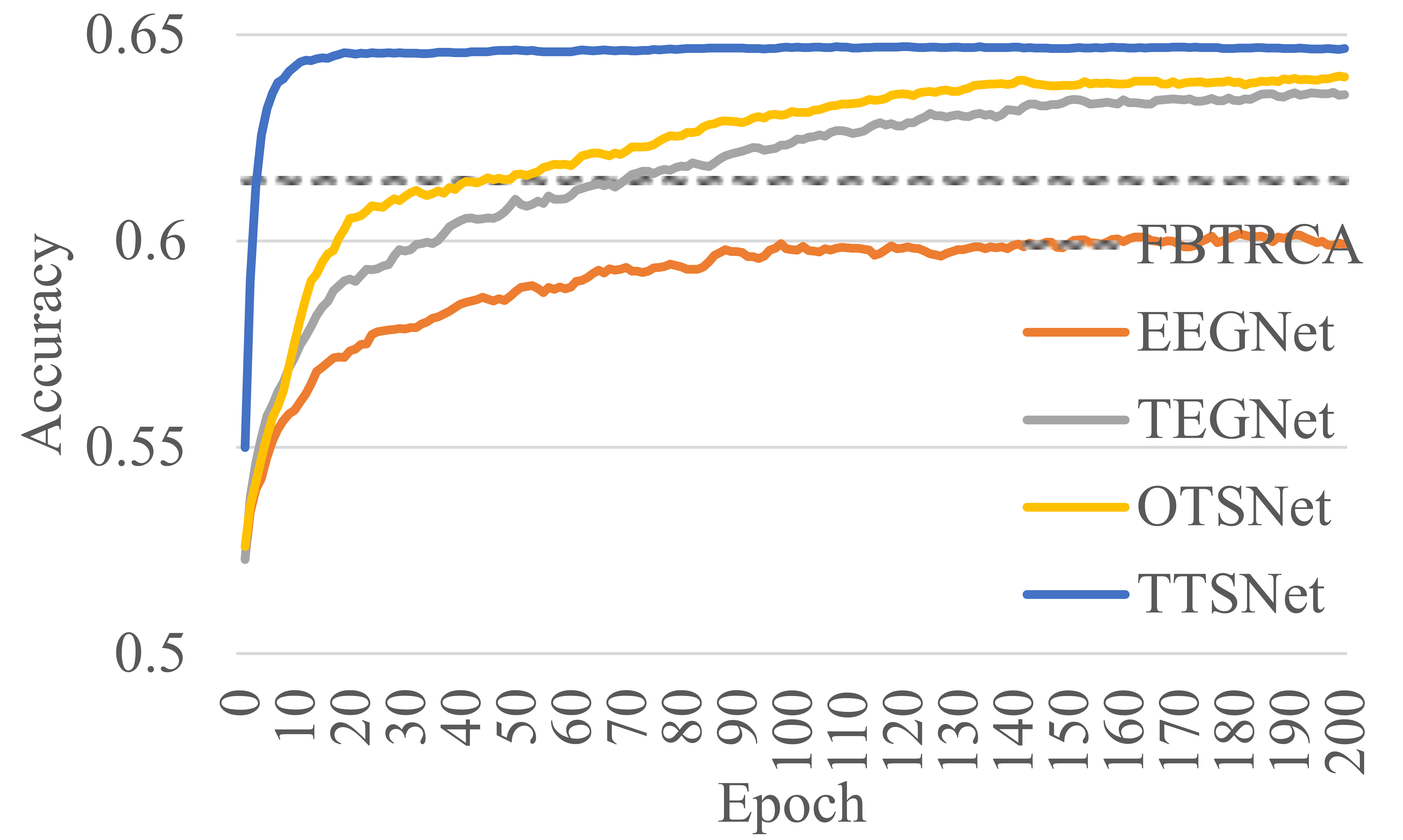}
    }
    \subfigure[Dataset II]{
        \includegraphics[width=0.315\textwidth]{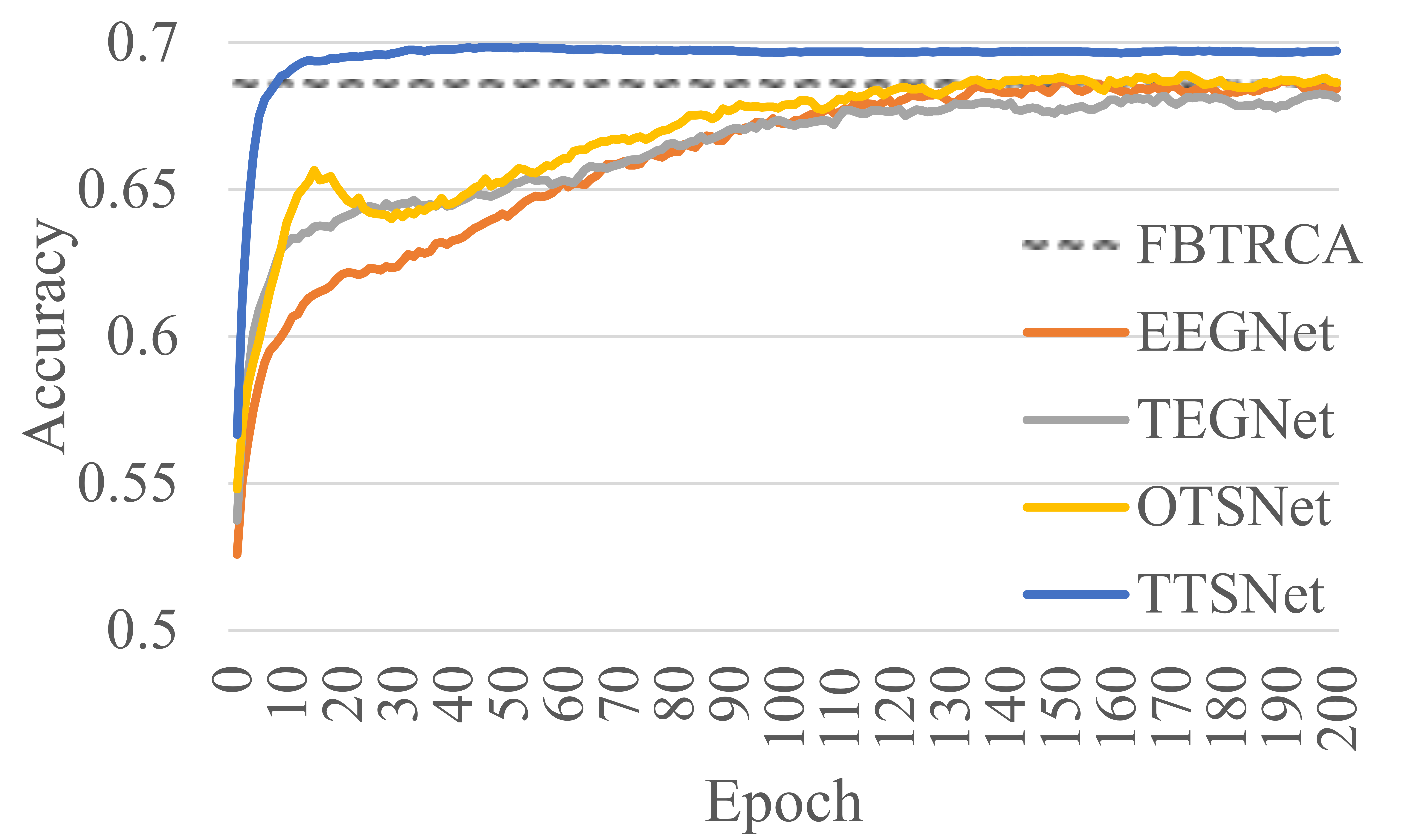}
    }
    \caption{Overall accuracy comparison in the binary classification.}
    \label{fig:3.1}
\end{figure*}

\begin{figure*}[!htbp]
    \centering
    \subfigure[Dataset I(a)]{
        \includegraphics[width=0.315\textwidth]{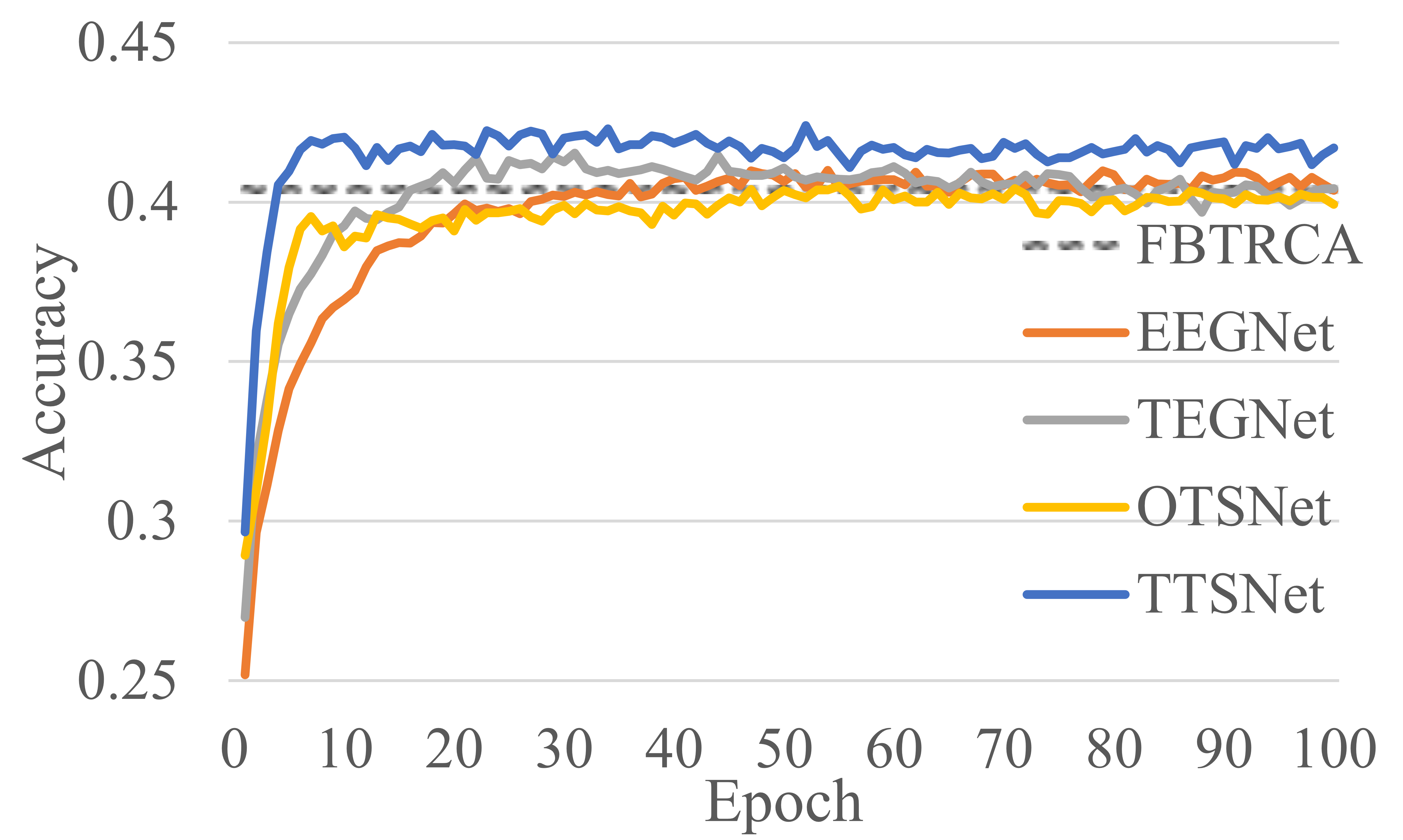}
    }
    \subfigure[Dataset I(b)]{
        \includegraphics[width=0.315\textwidth]{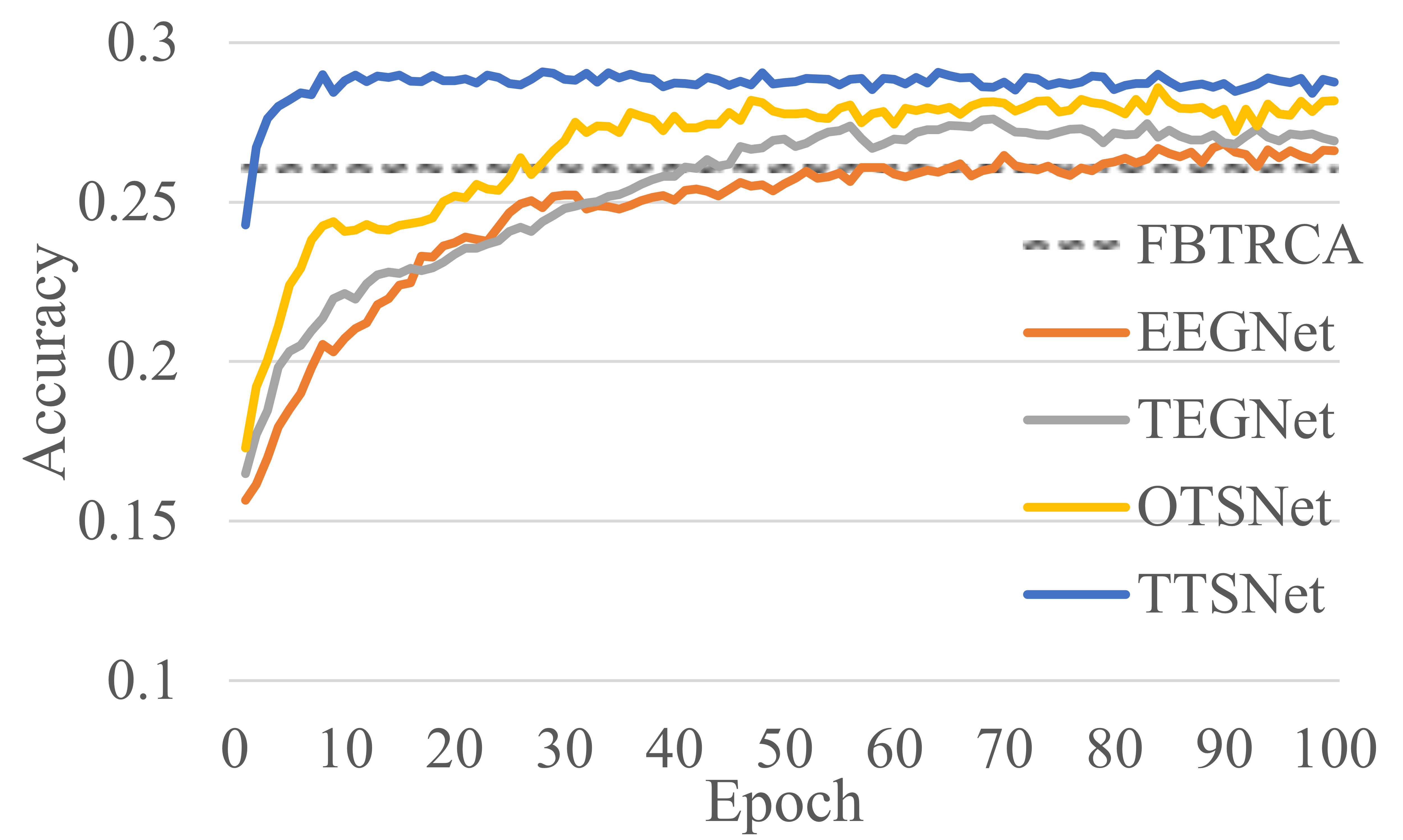}
    }
    \subfigure[Dataset II]{
        \includegraphics[width=0.315\textwidth]{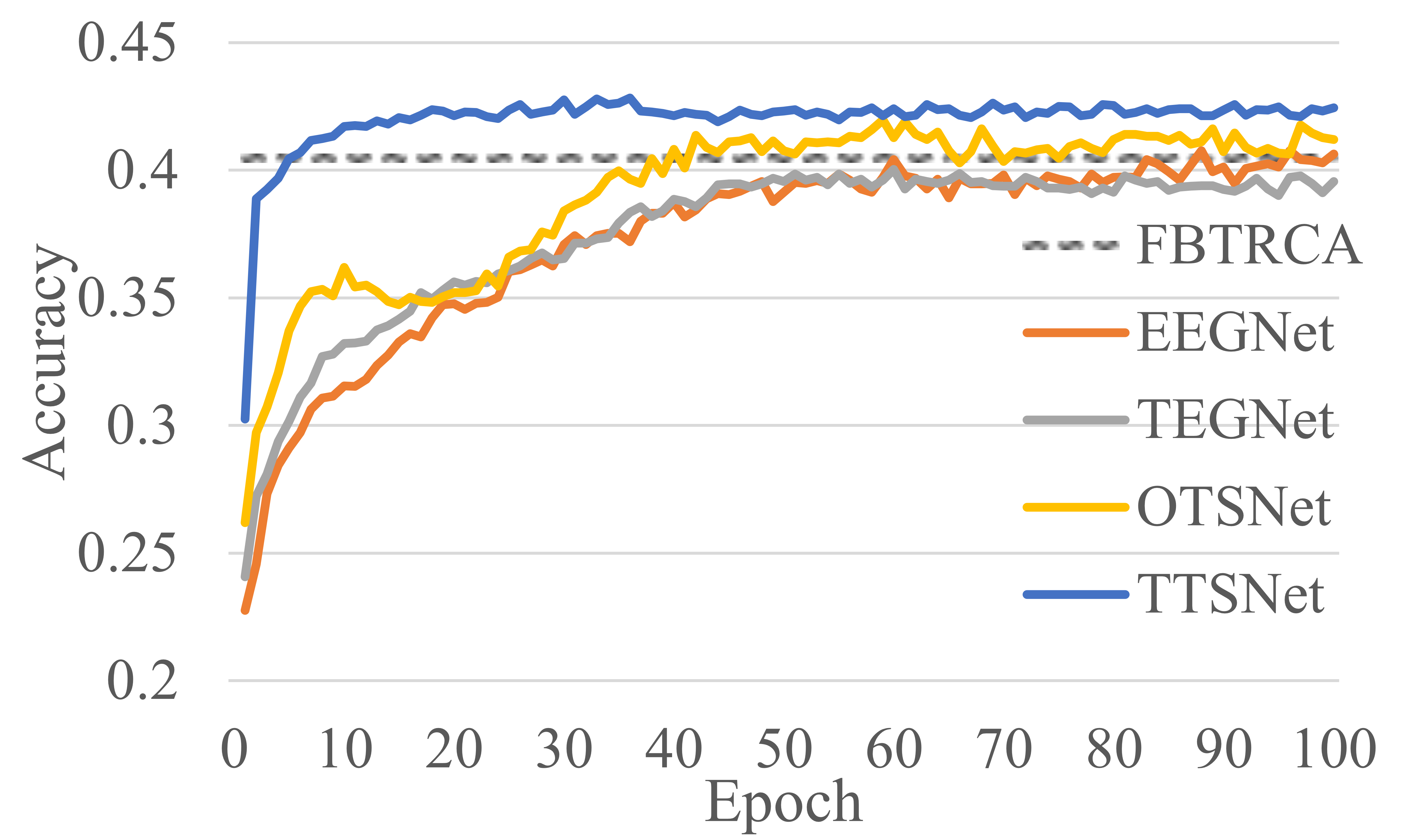}
    }
    \caption{Overall accuracy comparison in the multi-class classification.}
    \label{fig:3.2}
\end{figure*}

We compare the proposed method and the baseline methods by averaging the results of all subjects in each dataset. Figure \ref{fig:3.1} presents the average classification accuracy of each method in three cases in the binary classification. Figure \ref{fig:3.2} is the results as Figure \ref{fig:3.1} but in the multi-class classification. Table \ref{tab:3.1} summarizes the average accuracy of Figure \ref{fig:3.1} and Figure \ref{fig:3.2}. The accuracies given in this table are the accuracy after 200 training epochs in the binary classification and after 100 training epochs in the multi-class classification.

\begin{table*}[!htbp]
    \centering
    \caption{Overall average accuracy comparison}
    \label{tab:3.1}
    \begin{tabular}[]{c|c|c|c|c|c|c}
        \toprule
        \multirow{3}{*}{Dataset} & \multicolumn{2}{c}{Dataset I(a)} & \multicolumn{2}{c}{Dataset I(b)} & \multicolumn{2}{c}{Dataset II} \\
        \cmidrule{2-7}
         & Binary& Multi-class& Binary& Multi-class& Binary& Multi-class\\
        \midrule
        FBTRCA &0.7560$\pm$0.1190&0.4039$\pm$0.0745&0.6146$\pm$0.1290&0.2605$\pm$0.0561&0.6761$\pm$0.1136&0.4048$\pm$0.0742\\
        EEGNet &0.7213$\pm$0.1248&0.4036$\pm$0.0729&0.5991$\pm$0.1335&0.2660$\pm$0.0645&0.6843$\pm$0.1155&0.4063$\pm$0.0744\\
        TEGNet &0.7504$\pm$0.1164&0.4043$\pm$0.0726&0.6354$\pm$0.1318&0.2692$\pm$0.0612&0.6811$\pm$0.1142&0.3956$\pm$0.0773\\
        OTSNet &0.7545$\pm$0.1173&0.3992$\pm$0.0692&0.6398$\pm$0.1294&0.2817$\pm$0.0612&0.6863$\pm$0.1142&0.4121$\pm$0.0751\\
        TTSNet &\textbf{0.7619}$\pm$\textbf{0.1169}&\textbf{0.4170}$\pm$\textbf{0.0644}&\textbf{0.6466}$\pm$\textbf{0.1290}&\textbf{0.2876}$\pm$\textbf{0.0650}&\textbf{0.6972}$\pm$\textbf{0.1135}&\textbf{0.4244}$\pm$\textbf{0.0732}\\
        \bottomrule
    \end{tabular}
\end{table*}

\paragraph{Binary Classification}
Dataset I(a) shows that TEGNet has a better performance than EEGNet. It means that the spatial filter task-related component analysis used in the TEGNet can improve the performance of EEGNet. Meanwhile, the spatial filter minimizes the dimension of the input EEG signals and thus reduces the computation load in the training of EEGNet. In Dataset I(b) and Dataset II, the proposed TTSNet has improved performance compared to other methods. In dataset I(b), OTSNet and TEGNet have close performance and are better than the performance of FBTRCA and EEGNet.

\paragraph{Multi-class Classification}
In Dataset I(a), all the compared methods, except TTSNet, have a close average accuracy at about 40\%. The average accuracy of TTSNet is approximately 42\% and has an improvement of 2\%. In Dataset I(b) and Dataset II, OTSNet is better than FBTRCA, EEGNet, and TEGNet, but not as good as the TTSNet.

Relying on the overall analysis results, it shows that TTSNet has better performance than other methods when the movement onset cannot be located (Dataset I(a) and Dataset II). In the following paragraphs, we focus on the comparison of these methods in each subject of Dataset I(b) and Dataset II.

\subsubsection{Binary Performance within Motion Pairs}
We first present the classification performance in binary classification. In Dataset I(b), there are six movement states and the resting state. Table \ref{tab:3.2} lists the accuracies of binary classification between two movement states of Dataset I(b). Table \ref{tab:3.3} lists the accuracies between one movement state and the resting state of Dataset I(b). In Dataset II, there are five movement states. Table \ref{tab:3.4} is the accuracies between two motions in Dataset II. Three tables use the acronyms of the movement states and the resting state. For example, RE is the acronym of \textit{resting} state, and EE is that of \textit{elbow extension}.

Table \ref{tab:3.2}, \ref{tab:3.3}, and \ref{tab:3.4} show that the TTSNet has better performance than FBTRCA and EEGNet. Especially in the binary classification between one movement state and the resting state (Table \ref{tab:3.3}), TTSNet has an average accuracy of 0.7456$\pm$0.1205 across six motion pairs, which is greater than that of FBTRCA (0.6787$\pm$0.1260) and EEGNet (0.6506$\pm$0.1275).

\begin{table*}[htbp]
    \centering
    \scriptsize
    \caption{Binary Classification between Motion Pairs in Dataset I(b): Movement vs Movement}
    \label{tab:3.2}
    \begin{tabular}{|c|c|c|c|c|c|c|c|c|}
        \toprule
        \multirow{2}{*}{Method} & \multicolumn{8}{c|}{Motion Pair}\\
        \cmidrule{2-9}
        & EF-EE & EF-SU & EF-PR & EF-HC & EF-HO & EE-SU & EE-PR & EE-HC \\
        \midrule
        FBTRCA & 0.5317$\pm$0.1442  & 0.5728$\pm$0.1316  & 0.5750$\pm$0.1249  & 0.6078$\pm$0.1178  & 0.5928$\pm$0.1282  & 0.5650$\pm$0.1288  & 0.5756$\pm$0.1315  & 0.6156$\pm$0.1155 \\
        
        EEGNet & \textbf{0.5361}$\pm$\textbf{0.1287}  & 0.5600$\pm$0.1399  & 0.5628$\pm$0.1407  & 0.5972$\pm$0.1295  & 0.5744$\pm$0.1408  & 0.5350$\pm$0.1371  & 0.5461$\pm$0.1391  & 0.6261$\pm$0.1348 \\
        
        TEGNet & 0.5239$\pm$0.1477  & 0.5672$\pm$0.1467  & \textbf{0.6044}$\pm$\textbf{0.1358}  & 0.6139$\pm$0.1391  & 0.5789$\pm$0.1306  & \textbf{0.5922}$\pm$\textbf{0.1250}  & 0.6061$\pm$0.1269  & 0.6367$\pm$0.1320  \\
        
        OTSNet & 0.5339$\pm$0.1465  & \textbf{0.6061}$\pm$\textbf{0.1288}  & 0.5906$\pm$0.1299  & 0.6256$\pm$0.1363  & 0.5911$\pm$0.1354  & 0.5833$\pm$0.1386  & 0.5994$\pm$0.1342  & 0.6194$\pm$0.1169  \\
        
        TTSNet & 0.5211$\pm$0.1488  & 0.5950$\pm$0.1299  & 0.5972$\pm$0.1319  & \textbf{0.6383}$\pm$\textbf{0.1344}  & \textbf{0.5967}$\pm$\textbf{0.1278}  & 0.5878$\pm$0.1327  & \textbf{0.6094}$\pm$\textbf{0.1205}  & \textbf{0.6406}$\pm$\textbf{0.1371}  \\
        
        \midrule
        & EE-HO & SU-PR & SU-HC & SU-HO & PR-HC & PR-HO & HC-HO &\\
        \cmidrule{1-8}
        FBTRCA & 0.6006$\pm$0.1334  & 0.5456$\pm$0.1355  & 0.6200$\pm$0.1270  & 0.6350$\pm$0.1283  & 0.6283$\pm$0.1361  & 0.5939$\pm$0.1278  & \textbf{0.5756}$\pm$\textbf{0.1418}  &\\
        EEGNet & 0.5917$\pm$0.1465  & 0.5383$\pm$0.1389  & 0.6206$\pm$0.1349   & 0.6056$\pm$0.1382  & 0.6194$\pm$0.1271  & 0.6017$\pm$0.1338  & 0.5622$\pm$0.1274  &\\
        TEGNet & 0.6078$\pm$0.1342  & 0.5578$\pm$0.1257  & 0.6556$\pm$0.1303  & 0.6289$\pm$0.1476  & 0.6439$\pm$0.1394  & 0.6106$\pm$0.1444  & 0.5550$\pm$0.1305  &\\
        OTSNet & 0.6150$\pm$0.1416  & \textbf{0.5633}$\pm$\textbf{0.1176}  & \textbf{0.6583}$\pm$\textbf{0.1155}  & 0.6422$\pm$0.1337  & 0.6400$\pm$0.1336  & \textbf{0.6389}$\pm$\textbf{0.1249}  & 0.5650$\pm$0.1246 & \\
        TTSNet & \textbf{0.6156}$\pm$\textbf{0.1498}  & 0.5578$\pm$0.1318  & 0.6522$\pm$0.1156  & \textbf{0.6467}$\pm$\textbf{0.1332}  & \textbf{0.6533}$\pm$\textbf{0.1321}  & 0.6217$\pm$0.1304  & 0.5722$\pm$0.1298 & \\
        \bottomrule
    \end{tabular}
\end{table*}

\begin{table*}[htbp]
    \centering
    \caption{Binary Classification between Motion Pairs in Dataset I(b): Movement vs Resting}
    \label{tab:3.3}
    \begin{tabular}{|c|c|c|c|c|c|c|}
        \toprule
        \multirow{2}{*}{Method} & \multicolumn{6}{c|}{Motion Pair}\\
        \cmidrule{2-7}
        & EF-RE & EE-RE & SU-RE & PR-RE & HC-RE & HO-RE \\
        \midrule
        FBTRCA & 0.6839$\pm$0.1288  & 0.6883$\pm$0.1283  & 0.7106$\pm$0.1216  & 0.7028$\pm$0.1258  & 0.6283$\pm$0.1256  & 0.6583$\pm$0.1258  \\ 
        EEGNet & 0.6556$\pm$0.1331  & 0.6494$\pm$0.1197  & 0.6589$\pm$0.1269  & 0.6756$\pm$0.1339  & 0.6233$\pm$0.1315  & 0.6411$\pm$0.1202  \\
        TEGNet & 0.7250$\pm$0.1295  & 0.7217$\pm$0.1233  & 0.7633$\pm$0.1184  & 0.7722$\pm$0.1080  & 0.6722$\pm$0.1295  & 0.7072$\pm$0.1242  \\
        OTSNet & 0.7228$\pm$0.1208  & 0.7339$\pm$0.1214  & 0.7617$\pm$0.1231  & 0.7739$\pm$0.1151  & 0.6711$\pm$0.1385  & 0.6994$\pm$0.1408  \\
        TTSNet & \textbf{0.7483}$\pm$\textbf{0.1242}  & \textbf{0.7411}$\pm$\textbf{0.1184}  & \textbf{0.7800}$\pm$\textbf{0.1073}  & \textbf{0.7967}$\pm$\textbf{0.1160}  & \textbf{0.6872}$\pm$\textbf{0.1299}  & \textbf{0.7200}$\pm$\textbf{0.1273}  \\
        \bottomrule
    \end{tabular}
\end{table*}

\begin{table*}[htbp]
    \centering
    \scriptsize
    \caption{Binary Classification between Motion Pairs in Dataset II}
    \label{tab:3.4}
    \resizebox{\textwidth}{12mm}{
    \begin{tabular}{|c|c|c|c|c|c|c|c|c|c|c|}
        \toprule
        \multirow{2}{*}{Method} & \multicolumn{10}{c|}{Motion Pair}\\
        \cmidrule{2-11}
        & SU-PR & SU-HO & SU-PG & SU-LG & PR-HO & PR-PG & PR-LG & HO-PG & HO-LG & PG-LG \\
        \midrule
        FBTRCA & 0.6095$\pm$0.1084  & \textbf{0.6913}$\pm$\textbf{0.1145}  & 0.7325$\pm$0.1019  & 0.7079$\pm$0.1027  & 0.6817$\pm$0.1227  & 0.6889$\pm$0.1142  & 0.6865$\pm$0.1070  & 0.6833$\pm$0.1134  & 0.6587$\pm$0.1204  & 0.6206$\pm$0.1304  \\
        EEGNet & 0.5952$\pm$0.1288  & 0.6905$\pm$0.1239  & 0.7254$\pm$0.1016  & 0.7206$\pm$0.1015  & 0.6913$\pm$0.1230  & 0.7111$\pm$0.1131  & 0.6992$\pm$0.1125  & 0.6873$\pm$0.1076  & 0.6865$\pm$0.1120  & 0.6357$\pm$0.1315  \\
        TEGNet & 0.5881$\pm$0.1231  & 0.6746$\pm$0.1109  & 0.7302$\pm$0.1050  & 0.7325$\pm$0.1317  & 0.6921$\pm$0.1101  & 0.6857$\pm$0.1152  & 0.7175$\pm$0.1033  & 0.6913$\pm$0.1167  & 0.6667$\pm$0.1246  & 0.6325$\pm$0.1357  \\
        OTSNet & 0.6087$\pm$0.1295  & 0.6571$\pm$0.1074  & 0.7381$\pm$0.0994  & 0.7444$\pm$0.1048  & \textbf{0.7056}$\pm$\textbf{0.1046}  & 0.6960$\pm$0.1229  & 0.7087$\pm$0.1116  & 0.6921$\pm$0.1110  & 0.6794$\pm$0.1231  & 0.6333$\pm$0.1277  \\
        TTSNet & \textbf{0.6310}$\pm$\textbf{0.1259}  & 0.6770$\pm$0.1063  & \textbf{0.7452}$\pm$\textbf{0.1110}  & \textbf{0.7468}$\pm$\textbf{0.1127}  & 0.6976$\pm$0.1101  & \textbf{0.7119}$\pm$\textbf{0.1088}  & \textbf{0.7310}$\pm$\textbf{0.0978}  & \textbf{0.7024}$\pm$\textbf{0.1093}  & \textbf{0.6937}$\pm$\textbf{0.1260}  & \textbf{0.6357}$\pm$\textbf{0.1268}  \\
        \bottomrule
    \end{tabular}
    }
\end{table*}

\subsubsection{Multi-class Performance for Individual Subject} The multi-class classification performance for each of all the subjects is given in this part. Table \ref{tab:3.5} lists the accuracies of Dataset I(b), and Table \ref{tab:3.6} lists that of Dataset II.

\begin{table*}[!htbp]
    \centering
    \scriptsize
    \caption{Multi-class Classification Among All Motions in Dataset I(b)}
    \label{tab:3.5}
    \begin{tabular}{|c|c|c|c|c|c|c|c|c|}
        \toprule
        \multirow{2}{*}{Method} & \multicolumn{8}{c|}{Subject}\\
        \cmidrule{2-9}
        & 1 & 2 & 3 & 4 & 5 & 6 & 7 & 8 \\
        \midrule
        FBTRCA & 0.2524$\pm$0.0647  & 0.2524$\pm$0.0376  & 0.5905$\pm$0.0681  & 0.3500$\pm$0.0615  & 0.2286$\pm$0.0408  & 0.2762$\pm$0.0541  & 0.1810$\pm$0.0552  & 0.2214$\pm$0.0338  \\
        
        EEGNet & 0.2714$\pm$0.0746  & 0.2476$\pm$0.0729  & 0.6286$\pm$0.1011  & 0.3071$\pm$0.0520  & 0.2095$\pm$0.0672  & 0.2619$\pm$0.0817  & 0.1833$\pm$0.0503  & 0.2357$\pm$0.0482  \\
        
        TEGNet & \textbf{0.2714}$\pm$\textbf{0.0585}  & 0.2714$\pm$0.0340  & 0.6262$\pm$0.0753  & 0.2905$\pm$0.0474  & 0.2310$\pm$0.0539  & 0.2738$\pm$0.0780  & 0.2071$\pm$0.0736  & 0.2071$\pm$0.0594  \\
        
        OTSNet & 0.2500$\pm$0.0676  & 0.2714$\pm$0.0517  & 0.6690$\pm$0.0495  & 0.2952$\pm$0.0795  & \textbf{0.2333}$\pm$\textbf{0.0460}  & \textbf{0.3238}$\pm$\textbf{0.0834}  & 0.1976$\pm$0.0435  & \textbf{0.2381}$\pm$\textbf{0.0664}  \\
        
        TTSNet & 0.2571$\pm$0.0633  & \textbf{0.2929}$\pm$\textbf{0.0503}  & \textbf{0.6952}$\pm$\textbf{0.0784}  & \textbf{0.3476}$\pm$\textbf{0.0763}  & 0.2238$\pm$0.0694  & 0.2905$\pm$0.0726  & \textbf{0.2238}$\pm$\textbf{0.0927}  & 0.2310$\pm$0.0421  \\
        
        \midrule
        & 9 & 10 & 11 & 12 & 13 & 14 & 15 & \\
        \cmidrule{1-8}
        FBTRCA & 0.2333$\pm$0.0699  & 0.2476$\pm$0.0856  & 0.1881$\pm$0.0482  & 0.2000$\pm$0.0340  & 0.2262$\pm$0.0772  & 0.2262$\pm$0.0541  & 0.2333$\pm$0.0570  & \\
        EEGNet & 0.2310$\pm$0.0655  & 0.2952$\pm$0.0607  & 0.1929$\pm$0.0765  & 0.2167$\pm$0.0508  & 0.2310$\pm$0.0490  & 0.2357$\pm$0.0629  & 0.2429$\pm$0.0536  & \\
        TEGNet & 0.2357$\pm$0.0482  & 0.2690$\pm$0.0753  & 0.2429$\pm$0.0808  & 0.2024$\pm$0.0738  & 0.2595$\pm$0.0773  & 0.2119$\pm$0.0396  & 0.2381$\pm$0.0420  & \\
        OTSNet & 0.2238$\pm$0.0517  & 0.2976$\pm$0.0934  & \textbf{0.2452}$\pm$\textbf{0.0877}  & 0.2048$\pm$0.0376  & \textbf{0.2595}$\pm$\textbf{0.0495}  & 0.2524$\pm$0.0694  & 0.2643$\pm$0.0363  & \\
        TTSNet & \textbf{0.2357}$\pm$\textbf{0.0345}  & \textbf{0.3024}$\pm$\textbf{0.0664}  & 0.2048$\pm$0.0563  & \textbf{0.2190}$\pm$\textbf{0.0681}  & 0.2571$\pm$0.0751  & \textbf{0.2548}$\pm$\textbf{0.0770}  & \textbf{0.2786}$\pm$\textbf{0.0527} & \\
        \bottomrule
    \end{tabular}
\end{table*}

\begin{table*}[!htbp]
    \centering
    \scriptsize
    \caption{Multi-class Classification Among All Motions in Dataset II}
    \label{tab:3.6}
    \resizebox{\textwidth}{12mm}{
        \begin{tabular}{|c|c|c|c|c|c|c|c|c|c|}
            \toprule
            \multirow{2}{*}{Method} & \multicolumn{9}{c|}{Subject}\\
            \cmidrule{2-10}
            & 1 & 2 & 3 & 4 & 5 & 6 & 7 & 8 & 9\\
            \midrule
            FBTRCA & 0.3029$\pm$0.0788  & 0.3571$\pm$0.0526  & \textbf{0.3943}$\pm$\textbf{0.0735}  & 0.5429$\pm$0.0700  & 0.5057$\pm$0.0873  & 0.4371$\pm$0.0894  & 0.2714$\pm$0.0649  & 0.4343$\pm$0.0921  & 0.3971$\pm$0.0594  \\
            EEGNet & 0.3029$\pm$0.0728  & 0.3600$\pm$0.0689  & 0.3886$\pm$0.0964  & 0.5343$\pm$0.0572  & 0.5543$\pm$0.0663  & 0.4343$\pm$0.0710  & 0.2457$\pm$0.0865  & 0.4457$\pm$0.0663  & 0.3914$\pm$0.0842  \\
            TEGNet & 0.2914$\pm$0.0376  & 0.3286$\pm$0.0788  & 0.3743$\pm$0.0706  & 0.4943$\pm$0.0738  & 0.5429$\pm$0.0933  & 0.4143$\pm$0.0854  & 0.2686$\pm$0.0728  & 0.4286$\pm$0.0873  & 0.4171$\pm$0.0964  \\
            OTSNet & 0.3143$\pm$0.0819  & 0.3514$\pm$0.0863  & 0.3829$\pm$0.0715  & 0.5629$\pm$0.0572  & 0.5171$\pm$0.0579  & 0.4086$\pm$0.0894  & \textbf{0.3029}$\pm$\textbf{0.0776}  & \textbf{0.4486}$\pm$\textbf{0.0687}  & 0.4200$\pm$0.0852  \\
            TTSNet & \textbf{0.3171}$\pm$\textbf{0.0908}  & \textbf{0.3600}$\pm$\textbf{0.0335}  & 0.3800$\pm$0.0556  & \textbf{0.5657}$\pm$\textbf{0.0860}  & \textbf{0.5743}$\pm$\textbf{0.0857}  & \textbf{0.4543}$\pm$\textbf{0.0813}  & 0.2971$\pm$0.0715  & 0.4457$\pm$0.0740  & \textbf{0.4257}$\pm$\textbf{0.0802} \\
            \bottomrule
        \end{tabular}
    }
\end{table*}

Summarizing the results in this section, the improvement of TTSNet to FBTRCA and EEGNet has two different cases. When the movement onset can be located, TTSNet has a slight increase from 0.7560$\pm$0.1190 and 0.7213$\pm$0.1248 to 0.7619$\pm$0.1169. When the movement onset cannot be located, TTSNet has better performance than FBTRCA and EEGNet, with an improvement of 4.75\% in Dataset I and 2.11\% in Dataset II. Especially in the binary classification between one movement state and the resting state, TTSNet shows the improvement from 0.6787$\pm$0.1260 and 0.6506$\pm$0.1275 to 0.7456$\pm$0.1205, with 9.50\% increase.

\section{Discussion}
\label{sec:discuss}

In this work, we propose the TTSNet method, which combines our previous work FBTRCA and the EEGNet method. The spectral information from FBTRCA and the temporal information from EEGNet are fused in the TTSNet method. There are three key points in TTSNet, (1) filter bank, (2) spatial filter, and (3) two-stage training process.

Filter bank selection on MRCP signal processing is first used in our previous work on FBTRCA. When we developed the TTSNet, we noticed that EEGNet has its own drawback in signal processing. In EEGNet, the first two convolution layers are used to preprocess the raw signals and extract the necessary features, and then the following layers are used to compress and select the essential features (see Table \ref{tab:1}). In EEG signal processing, the analysis in the frequency domain, temporal domain, and spatial domain is the universal solution. In the preprocessing of raw signals in EEGNet, the first convolution layer is equipped with the 1$\times$64 kernel size. The second layer is equipped with the grouped $C$$\times$1 kernel size, where $C$ is the number of channels. The two layers are related to the feature extraction from the temporal domain and spatial domain, respectively. However, the analysis in the frequency domain is ignored in EEGNet. Therefore, we assume that the EEGNet equipped with filter bank selection can further improve the performance of EEGNet. The results in this work verify this assumption because TTSNet has an increased classification accuracy than EEGNet.

The spatial filter, TRCA, is first used in the binary classification of two motions in MRCP signal processing. It is adjusted to the multi-class classification in the following study on STRCA. In this work, the spatial filter used is the adjusted version. TRCA aims to find a projection that maximizes the inter-trial covariance within a class. By maximizing the covariance, TRCA rejects the unrelated signals from the raw EEG signals. In this work, we fuse this spatial filter and the EEGNet and get the TEGNet. There are two reasons to use TEGNet instead of EEGNet. First, the spatial filter can decrease the dimension of the input raw EEG signals and thus reduce the computation load of EEGNet. Second, TEGNet can improve the performance of EEGNet, either the convergence speed in the training of the network or the classification accuracies.

In the training process of the TTSNet, we adopt the two-stage training. The OTSNet and TTSNet have the same network structure. In the OTSNet, we construct the network and train the network directly. In the TTSNet, we first train the parameters of EEGNet for each of the filter banks. The parameters of EEGNet are fixed when the training process converges after a few epochs. The fully connected layer is then trained with the output of EEGNets of these filter banks in the second step. The two-stage training inherits the idea of FBTRCA. In FBTRCA, the canonical correlation patterns are extracted from the spatial-filtered EEG signals. The canonical correlation patterns and EEGNet play the same role in signal processing. Canonical correlation patterns measure the similarity between the single-trial EEG signal and the grand average MRCPs of multiple classes with two-dimensional correlation (Equation \ref{eqn:corr}). The grand average MRCP is the pre-trained signals by averaging the EEG trials belonging to the same class. For each class, FBTRCA will compute three kinds of correlations. Correlations of all classes and filter banks are concatenated and used in the following feature selection. The EEGNet in the two-stage training process replaces the above calculation of the correlation. 

The FBTRCA and the TTSNet have the same framework, as in Figure \ref{fig:2.1}. This framework requires a two-stage training process. The spatial filter in this framework cannot discriminate the classes of the EEG signals but just reject the irrelevant noises and decrease the dimension of signals. Before the canonical correlation or the EEGNet, there was no operation that served for classification but for preprocessing. Filter bank division can be used to improve the classification accuracies in pre-movement decoding. However, the methods are supposed to work without filter bank divisions, such as the STRCA and the TEGNet.

The performance improvement from FBTRCA to TTSNet can be explained as follows. As mentioned above, the differences between FBTRCA and TTSNet are the extracted features. TTSNet replaces the canonical correlation patterns in FBTRCA with EEGNet. The correlation is the element-wise multiplication between the single-trial signals and the grand average MRCPs. The grand average MRCP can be regarded as a pre-trained weight, and the correlation computes the weighted single-trial signals. The grand average MRCP averages all the trials belonging to the same class so that it can smooth the signals and reduce the influence of random noises to some degree. However, the pre-trained weight in FBTRCA may not be accurate because of the unexpected outliers and noises. Especially in the case that the movement onset cannot be located, the pre-trained weight is heavily eliminated by inter-trial delay. The pre-trained weight is thus replaced with the kernels in the CNN model. The shift-invariant of the CNN kernel can reduce the negative influence of inter-trial delay to some degree. The kernel can be trained and adjusted during the training process, and a more stable and accurate weight can be obtained.

Although the proposed TTSNet uses information from the EEG signals in the temporal domain and the frequency domain, this method cannot fully use spatial information. The spatial filter, TRCA, functions as the noise rejection and dimension decrease. The more detailed information cannot be decoded with this spatial filter, such as the source location of the task-related components. Our future work will focus on the analysis of spatial information of the MRCP signals.

\section{Conclusion}
\label{sec:conclu}
In this work, we propose the TTSNet that uses temporal and spectral information to decode patterns from the MRCP signals. TTSNet incorporates the temporal decoding of EEGNet into the FBTRCA method and thus decodes the temporal and spectral information into distinct features. This method serves for both the binary and multi-class classification of upper limb movements. The results show that TTSNet has better classification performance than FBTRCA and EEGNet in the decoding of pre-movement patterns, especially in the case that the movement onset cannot be located. This work is expected to help the rehabilitation of the disabled or weak upper limbs. The code repository link to this work is \url{https://github.com/plustar/Movement-Related-Cortical-Potential.git}.

\section{Acknowledgement}
C.F.C work  was partially supported by grants PICT 2020-SERIEA-00457 and PIP 112202101 00284CO (Argentina).

\bibliographystyle{unsrt}
\bibliography{main}

\end{document}